\documentclass{appolb}

\usepackage{amsmath,amstext,amsthm,amssymb}
\usepackage{graphicx}
\usepackage{booktabs}

\newcommand{\ve}[1]{\ensuremath{\mathbf{#1}}}
\newcommand{\n}[1]{\ensuremath{|\mathbf{#1}|}}

\newcommand{\bref}[1]{Ref.~\cite{#1}}

\newcommand{\sref}[1]{Sec.~\ref{#1}}

\newcommand{\eref}[1]{Eq.~\eqref{#1}}
\newcommand{\Fref}[1]{Fig.~\ref{#1}}
\newcommand{\fref}[1]{Fig.~\ref{#1}}

\newcommand{\Figref}[1]{Figure~\ref{#1}}
\newcommand{\figref}[1]{Figure~\ref{#1}}

\begin{document}
\title{APPROXIMATIONS OF THE SPECTRAL FUNCTION\footnote{Presented at the XX Max Born Symposium ``Nuclear Effects in Neutrino Interactions'',
Wroc{\l}aw, Poland, December 7~–-10, 2005.}}
\author{ARTUR M. ANKOWSKI
\address{Institute of Theoretical Physics, University of Wroc{\l}aw,\\
pl. M.~Borna 9, 50-204 Wroc{\l}aw, Poland\\
{\tt artank@ift.uni.wroc.pl}}
}

\maketitle

\begin{abstract}
The ICARUS and future liquid argon neutrino experiments generate
demand for evaluating the spectral function of  argon. In this paper
we use oxygen nucleus as a~testing ground for our phenomenological
approach to the spectral function and probe the influence of
momentum distribution and treatment of the mean field spectral
function on the differential cross sections. The obtained model
reproduces very well results of the exact spectral function of
oxygen and can be applied to heavier nuclei, such as calcium or
argon.
\end{abstract}

\section{Motivation and outline of the paper}
Thanks to experience already gained with the ICARUS T600
TPC~\cite{ref:ICARUS} one knows that liquid argon (LAr) has many
advantages in neutrino experiments. They make LAr technology
interesting for planned detectors, e.g. for T2K~\cite{ref:T2K} and
NuMI~\cite{ref:NuMI}. To use them fully, one has to reduce
uncertainties by evaluating nuclear effects as precise as possible.
From ($e$, $e'$) scattering it's clear that the Fermi gas model can
only be the first approximation.

More elaborate approach is the spectral function (SF) formalism.
Where the impulse approximation is valid (i.e. when neutrino
energy~$E_\nu$ is greater then a~few hundred of
MeV~\cite{ref:Co_priv}), the SF describes nuclei most accurately. It
was applied to ($l$, $l'$) scattering previously~\cite{ref:Oset}.
The problem is that the exact SFs exist only for a~few double closed
shell nuclei, $^3$He, and nuclear matter
\cite{ref:Benhar&Fabrocini&Fantoni&Sick, ref:Benhar&Farina,
ref:Benhar&Pandharipande}. For argon~$^{40}_{18}$Ar exact
computations cannot be performed, so one is forced to seek for the
best possible approximation. Opportunity to verify the quality of
given approximation is provided by oxygen nucleus, where the exact
SF exist~\cite{ref:Benhar&Farina}: applying the same method to
calculation of the oxygen SF and comparing the result to the exact
SF can give a~notion of discrepancies between them.

The first attempt at constructing the argon SF is
\bref{ref:Ankowski&Sobczyk}. Presented there approximation gives the
differential cross sections which somewhat differ from the
corresponding ones for the exact SF (compare \fref{fig:sf}). In the
paper presumption was made that the discrepancies come from:
\begin{itemize}
\item[---]{oversimplified treatment of the mean field spectral function,}
\item[---]{different momentum distributions in the spectral functions.}
\end{itemize}
This work is devoted to detailed study of the two effects. After
a~brief introduction into the SF approach in \sref{sec:SFApproach}
and describing the simplest approximation of the SF in
\sref{sec:SSF} (i.e. the approach of~\cite{ref:Ankowski&Sobczyk}),
we consider in \sref{sec:MFTreatment} an influence of
NN-correlations on the mean field SF. Then in \sref{sec:MomDistrib}
the sensitivity of the SF on the momentum distribution is discussed.

Conclusions allowed for working out in \sref{sec:GSF} a~satisfactory
approximation of the exact spectral function of oxygen, which can be
applied to argon and other nuclei.

\section{Basic information on the spectral function}\label{sec:SFApproach}
The spectral function (SF) of a given nucleus $P(\ve p, E)$ is the
probability distribution of finding a~nucleon with momentum~$\ve p$
and removal energy $E$. Formal definition can be expressed
as~\cite{ref:Frullani&Mougey}
\begin{equation}\label{eq:DefOfSF}
P(\ve p,E)\equiv\langle i(M_{A})|a^{\dagger}(\ve p)\delta(\hat H-M_A+M-E)a(\ve p)|i(M_{A})\rangle,%
\end{equation}
where $\hat H$ is the intrinsic Hamiltonian of the $(A-1)$-nucleon
system, $|i(M_{A})\rangle$ denotes the state of the initial nucleus
of mass $M_A$ (assumed to be at rest), and $M$ is nucleon mass.

Oxygen nucleus is isoscalar ($N=Z$), therefore its SF consists of
only two parts: the mean field and the short-range correlation one:
\begin{equation}\label{eq:ContribsToSF}
P(\ve p,E)=\frac{N+Z}2\left[P_\text{MF}(\ve p,E)+P_\text{corr}(\ve p,E)\right].%
\end{equation}

All the knowledge about low-energy and low-momentum nucleons (i.e.
the shell model information) is included in the mean field term
$P_\text{MF}(\ve p,E)$. Corrections to the independent-particle
behavior of nucleons are described by the correlation part
$P_\text{corr}(\ve p,E)$. From the nuclear matter calculations one
knows that at high energy and momentum such correlations are
dominated by the two-nucleon
interactions~\cite{ref:Benhar&Fabrocini&Fantoni}.

The mean field SF consists of contributions of every shell model
state~$\alpha$ below the Fermi level
$\alpha_F$~\cite{ref:Ciofi&Simula}:
\begin{equation}\label{eq:MFSF}
P_\text{MF}(\ve p,E)\equiv\frac1A\sum_{\alpha<\alpha_F}c_\alpha A_{\alpha}|\phi_\alpha(\ve p)|^2\delta\big(E_{\alpha}+E_R(\ve p)-E\big).%
\end{equation}
In the above equation $E_\alpha$ is the energy of the state~$\alpha$
described by the single-nucleon wave function $\phi_\alpha(\ve p)$
(normalized to 1) with the occupation probability $c_\alpha$, and
the number of particles $A_\alpha$, whereas $E_R(\ve p)={\ve
p^2}/{(2M_{A-1})}$ is the recoil energy of the residual nucleus.

Interactions between nucleons cause their partial redistribution
from the states below the Fermi level to the levels of higher
energy, and occurrence of non-zero width of each level. It means
that when NN-correlations in the MF part are taken into account,
delta function in \eref{eq:MFSF} should be replaced by appropriate
distribution~\cite{ref:Benhar&Farina}. We shall return to this issue
in \sref{sec:MFTreatment}.

Due to \eref{eq:DefOfSF} the momentum distribution of nucleons reads
\begin{equation}
n(\ve p)\equiv\langle i(M_{A})|a^{\dagger}(\ve p)a(\ve p)|i(M_{A})\rangle=\int P(\ve p, E)\,d E%
\end{equation}
and as a consequence of \eref{eq:ContribsToSF} consists of two
contributions~\cite{ref:Ciofi&Simula}:
\begin{equation}
n(\ve p)=\frac{N+Z}2\left[n_\text{MF}(\ve p)+n_\text{corr}(\ve p)\right].%
\end{equation}

\section{Simplest approximation of the SF}\label{sec:SSF}
Firstly note that \eref{eq:MFSF} simplifies significantly when one
replaces $E_{\alpha}$ in the argument of the delta function by the
average separation energy $E^{(1)}=\sum_{\alpha<\alpha_F}c_\alpha
E_\alpha
 A_{\alpha}/A$~\cite{ref:Kulagin&Petti}:
\begin{eqnarray}\label{eq:MFSSF}
P^\text{SSF}_\text{MF}(\ve p,E)=n_\text{MF}(\ve p)\,\delta\big(E^{(1)}+E_R(\ve p)-E\big),%
\end{eqnarray}
because then only the momentum distribution
\begin{equation}\label{eq:nMF}
n_\text{MF}(\ve p)=\sum_{\alpha<\alpha_F}c_\alpha|\phi_\alpha(\ve p)|^2 A_{\alpha}/A%
\end{equation}
occurs, which is averaged over single-particle levels.

\begin{figure}
\graphicspath{{plot/}}\centering
\begin{minipage}[c]{0.49\textwidth}%
\includegraphics*[scale=0.48]{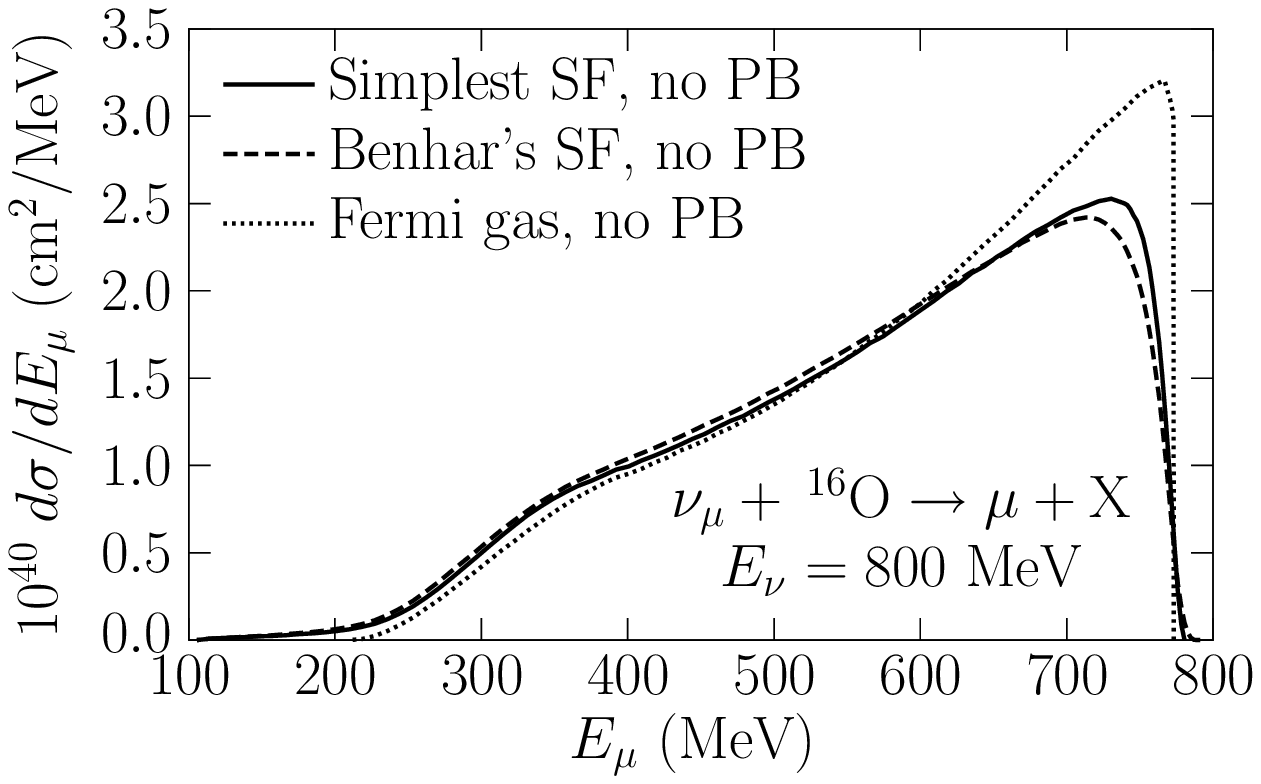}
\end{minipage}
\begin{minipage}[c]{0.49\textwidth}%
\includegraphics*[scale=0.48]{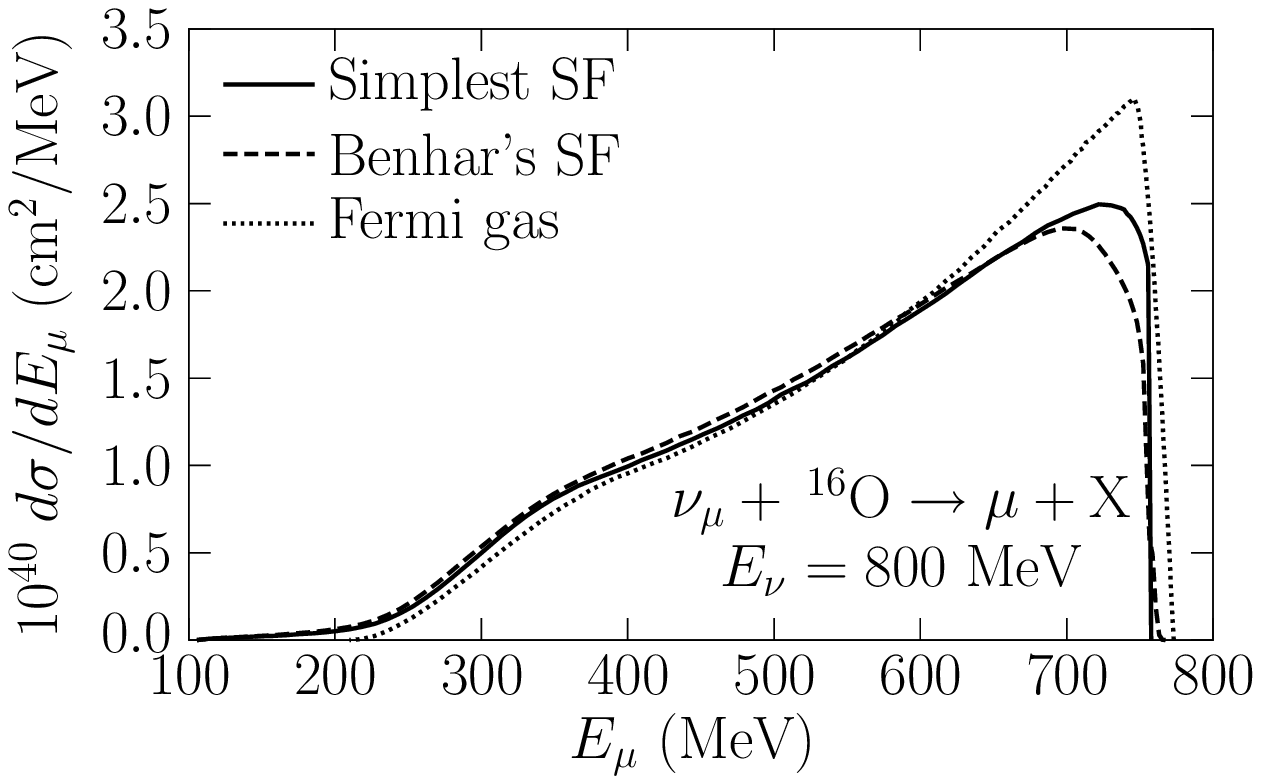}
\end{minipage}
\caption{Differential cross section $d\sigma/dE_\mu$ of quasielastic
$\nu_\mu$ scattering off $^{16}$O obtained from the Fermi gas model
(dotted line), the Benhar's spectral function (dashed line) and the
simplest approximation of the spectral
function~\cite{ref:Ankowski&Sobczyk} (solid line).\break Left: Pauli
blocking is absent. Right: Pauli blocking included.}\label{fig:sf}
\end{figure}

Secondly, as it was mentioned in the previous section, major
contribution to the correlations between nucleons comes from the
two-nucleon process. It consists in forming a~cluster by two
nucleons with high relative momentum while the other $(A-2)$
nucleons remain soft~\cite{ref:Ciofi&Liuti&Simula}. When we restrict
ourself to such an interactions, the correlation SF can be expressed
as~\cite{ref:Kulagin&Petti}:
\begin{equation}\label{eq:correlationSF}
P_\text{corr}(\ve p,E)=n_\text{corr}(\ve p)\frac{M}{\n p}\sqrt{\frac\alpha\pi}\left[\exp(-\alpha \ve p_\text{min}^2)-\exp(-\alpha\ve p_\text{max}^2)\right].%
\end{equation}
Occurring here $\alpha=3/(4\langle \ve p^2\rangle \beta)$ is
inversely proportional to the mean value of the MF momentum
squared~$\langle \ve p^2\rangle$ times $\beta=(A-2)/(A-1)$ and
\begin{equation}\begin{split}
{\ve p}_\text{min}^2&=\left[\beta \n p - \sqrt{\smash[b]{2M\beta[E-E^{(2)}-E_R(\ve p)]}}\,\right]^2,\\%
{\ve p}_\text{max}^2&=\left[\beta \n p + \sqrt{\smash[b]{2M\beta[E-E^{(2)}-E_R(\ve p)]}}\,\right]^2.%
\end{split}\end{equation}
The threshold value $E^{(2)}$ is interpreted as the two-nucleon
separation energy averaged over low-energy configurations of the
$(A-2)$-nucleon system and can be approximated by
$E^{(2)}=M_{A-2}+2M-M_A$.


We will refer to the described approach as ``the simplest
approximation of the spectral function'' or ``the simplest spectral
function'' (SSF).


\Figref{fig:sf} illustrates that the cross section $d\sigma/dE_\mu$
of both SFs clearly differs from the one of the Fermi gas. As long
as Pauli blocking (PB) is not included, the SSF reproduces the
result of the Benhar's (i.e. exact) SF quite well. Presence of PB
enlarges difference between them, because it changes the shape of
the peak in other way.

In the next section we investigate whether more elaborate
approximation of the mean field SF removes this discrepancy.


\section{Improved treatment of the mean field SF}\label{sec:MFTreatment}
In \sref{sec:SFApproach} we mentioned that NN-correlations broaden
energy levels, so in \eref{eq:MFSF} instead of delta function an
appropriate distribution should occur~\cite{ref:Benhar&Farina}:
\begin{equation}
P_\text{MF}(\ve p,E)\equiv\frac1A\sum_{\alpha<\alpha_F}c_\alpha A_{\alpha}|\phi_\alpha(\ve p)|^2F_\alpha\big(E_{\alpha}+E_R(\ve p)-E\big).%
\end{equation}
In order to simplify it, let's replace the single-particle momentum
distribution $|\phi_\alpha(\ve p)|^2$ by the average
$n_\text{MF}(\ve p)$ defined in \eref{eq:nMF}:
\begin{equation}
P_\text{MF}(\ve p,E)=n_\text{MF}(\ve p)\frac1A\sum_{\alpha<\alpha_F}c_\alpha A_{\alpha}F_\alpha(E_{\alpha}+E_R(\ve p)-E).%
\end{equation}

We decided to use the Gaussian distribution
\begin{equation}\label{eq:Gauss}
F_\alpha(x)=1/({\sqrt\pi D_\alpha})\exp\left[-\left({x/D_\alpha}\right)^2\right].%
\end{equation}

\begin{figure}
\graphicspath{{plot/}}\centering
\includegraphics*[scale=0.49]{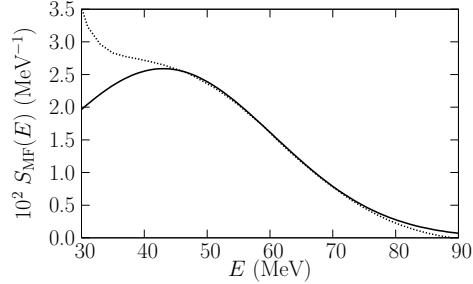}
\caption{Our fit of the Gaussian distribution of the level
$1s\frac12$ (solid line) to appropriate region of the energy
distribution $S_\text{MF}(E)$ calculated from the Benhar's spectral
function of $^{16}${O} (dotted line).}\label{fig:EBenDistrib}
\end{figure}

To describe a~level $\alpha$ we need to now its diffuseness
$D_\alpha$ and its ``mean field'' occupation probability $c_\alpha$.
If one used the ``raw'' $c_\alpha$'s (given for example
in~\cite{ref:OccProbs}), partial double counting would occur due to
contribution of the correlation term of the SF (raw value says only
that the nucleon can be found in given state with certain
probability, not how it get there: whether it happened due to
``natural placement'' or due to ``correlation kick'').

For $^{16}${O} we shall determine the value of $D_\alpha$ from the
Benhar's SF, for other nuclei one should get them from direct
calculations (e.g. for $^{40}_{20}${Ca} the data is available
in~\cite{ref:CaData}). Oxygen is a quite heavy nucleus in the sense
that it's recoil energy for the mean value of momentum $\n p=180$
MeV/$c$~\cite{ref:Ankowski&Sobczyk} is $\sim 1$ MeV. When we neglect
it, the separation
\begin{equation}
P_\text{MF}(\ve p,E)=n_\text{MF}(\ve p)S_\text{MF}(E).%
\end{equation}
holds, and we easily get that
\begin{equation}
S_\text{MF}(E)\propto \int P_\text{MF}(\ve p, E)\,d^3 p =\int \left[P(\ve p, E)-P_\text{corr}(\ve p, E)\right]\,d^3 p.%
\end{equation}
We inserted in the above relation the Benhar's SF and calculated the
energy distribution $S_\text{MF}(E)$. By fitting \eref{eq:Gauss}
successively to every peak corresponding to the energy level, we
determined all $D_\alpha$'s, see \figref{fig:EBenDistrib}.

Note that $S_\text{MF}(E)$ shown in \fref{fig:EBenDistrib} vanishes
for $E\gtrsim 87$ MeV. For oxygen $P_\text{corr}(\ve p, E)$ appears
above the threshold energy $E^{(2)}=26.33~\text {MeV}$ higher then
the energy of quite sharp levels $1p\frac32$ and $1p\frac12$, hence
the correlations admixes only to $c_\alpha$ of $1s\frac12$ level.
After subtracting calculated value of $\int S_\text{corr}(E)dE$ over
$E\in[26.33;87]$ MeV from $c_\alpha$ of $1s\frac12$, we get the
proper ``mean field'' value. Now we have all the
in{\-}dis{\-}pen{\-}sable parameters to obtain predictions for the
model.

\begin{figure}
\graphicspath{{plot/}}\centering
\begin{minipage}[l]{0.49\textwidth}%
\includegraphics*[scale=0.48]{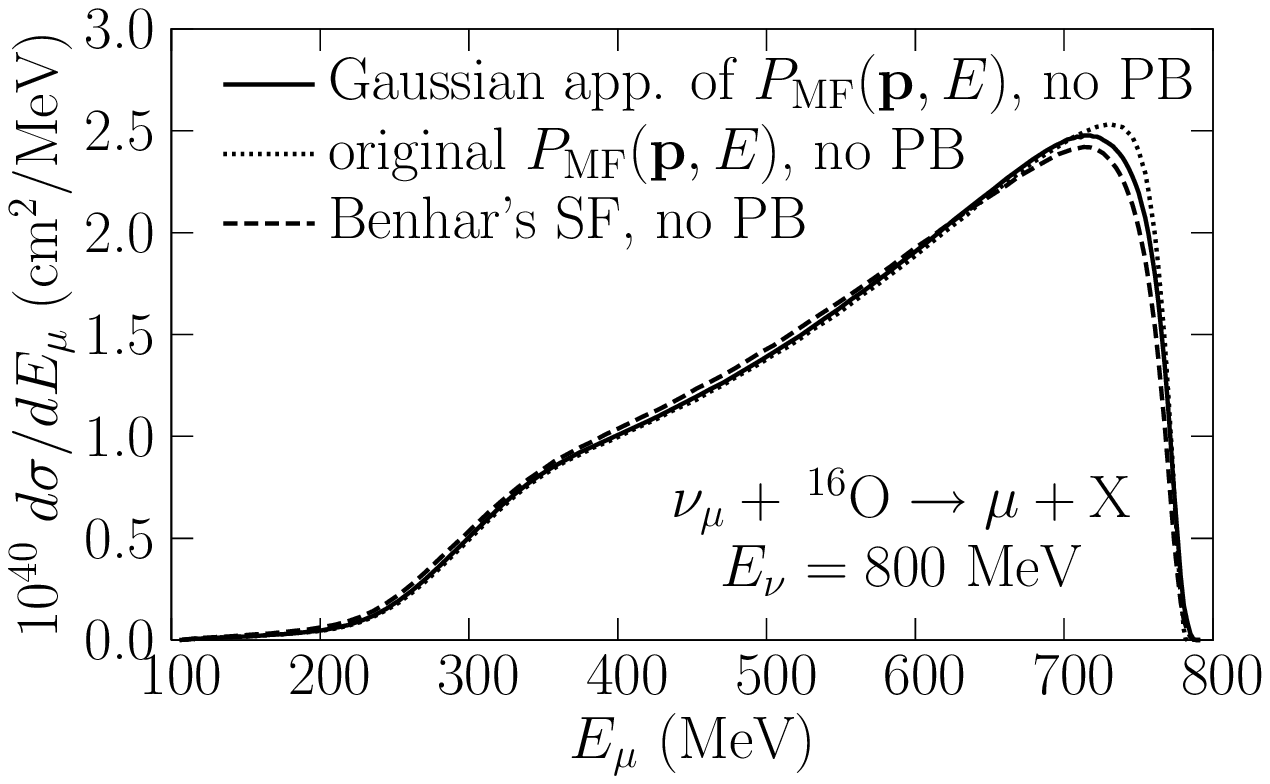}
\end{minipage}
\begin{minipage}[r]{0.49\textwidth}%
\includegraphics*[scale=0.48]{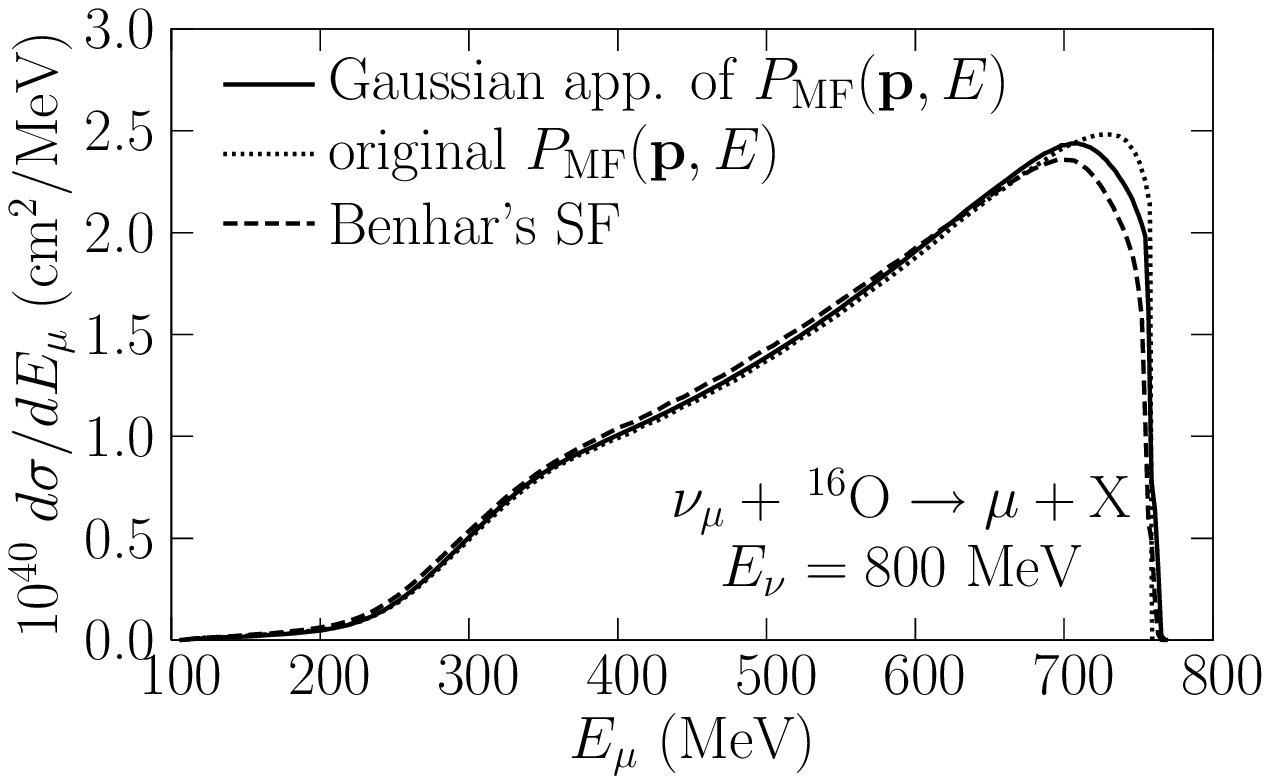}
\end{minipage}
\caption{Sensitivity of the quasielastic differential cross section
$d\sigma/dE_\mu$ on treatment of the mean field spectral function.
The simplest (dotted line) and the Gaussian approximation (solid
line) compared to the Benhar's spectral function. Left: Results
without Pauli blocking. Right: Pauli blocking included.
}\label{fig:cgsf}
\end{figure}

The cross sections for this approximation are shown in
\fref{fig:cgsf}: the momentum distribution is the same as in the
simplest SF and so is the correlation part of the SF, the only
change is the different treatment of the mean field SF. This
approach causes that the shape of the cross section is similar to
one of the exact SF, regardless of including PB or not. The height
of the peak is also slightly reduced, therefore the agreement
between the approximation and the Benhar's SF is now better.

\section{Dependence on the momentum distribution}\label{sec:MomDistrib}%

\begin{figure}
\graphicspath{{plot/}}\centering
\begin{minipage}[c]{0.49\textwidth}%
\includegraphics*[scale=0.49]{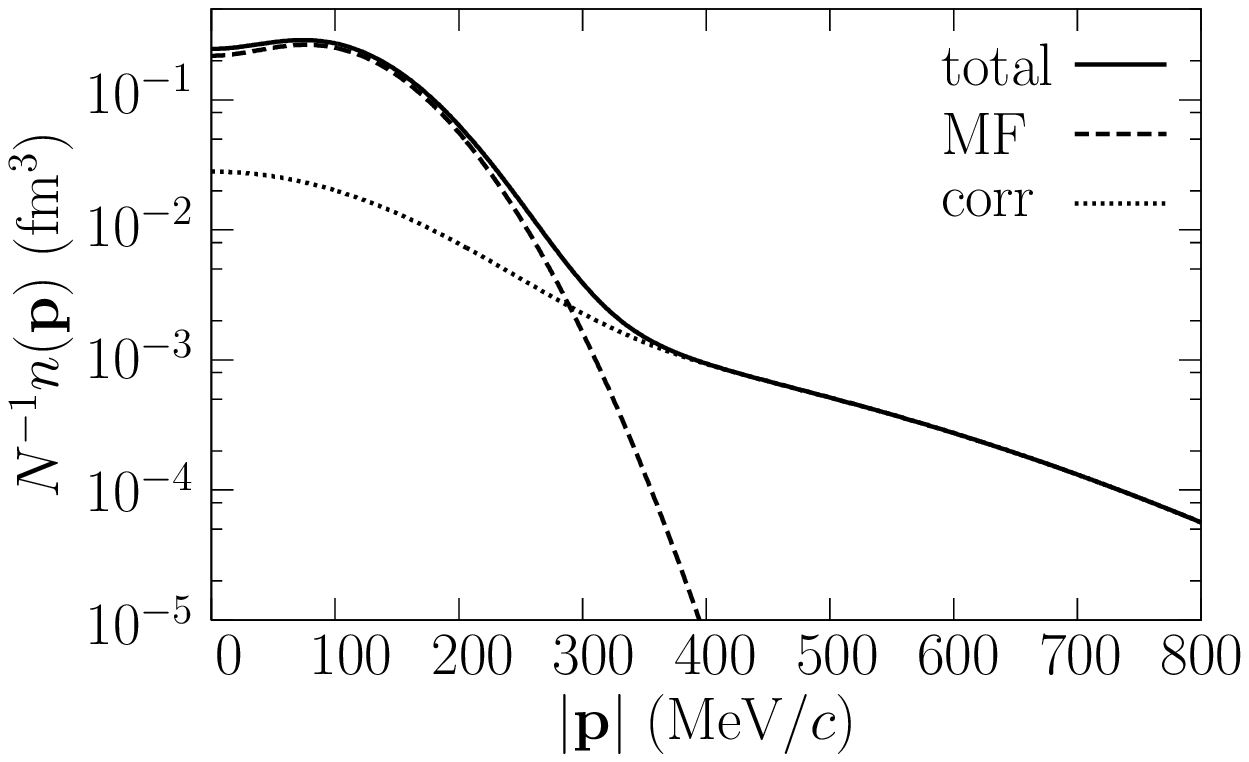}
\end{minipage}
\begin{minipage}[c]{0.49\textwidth}%
\includegraphics*[scale=0.49]{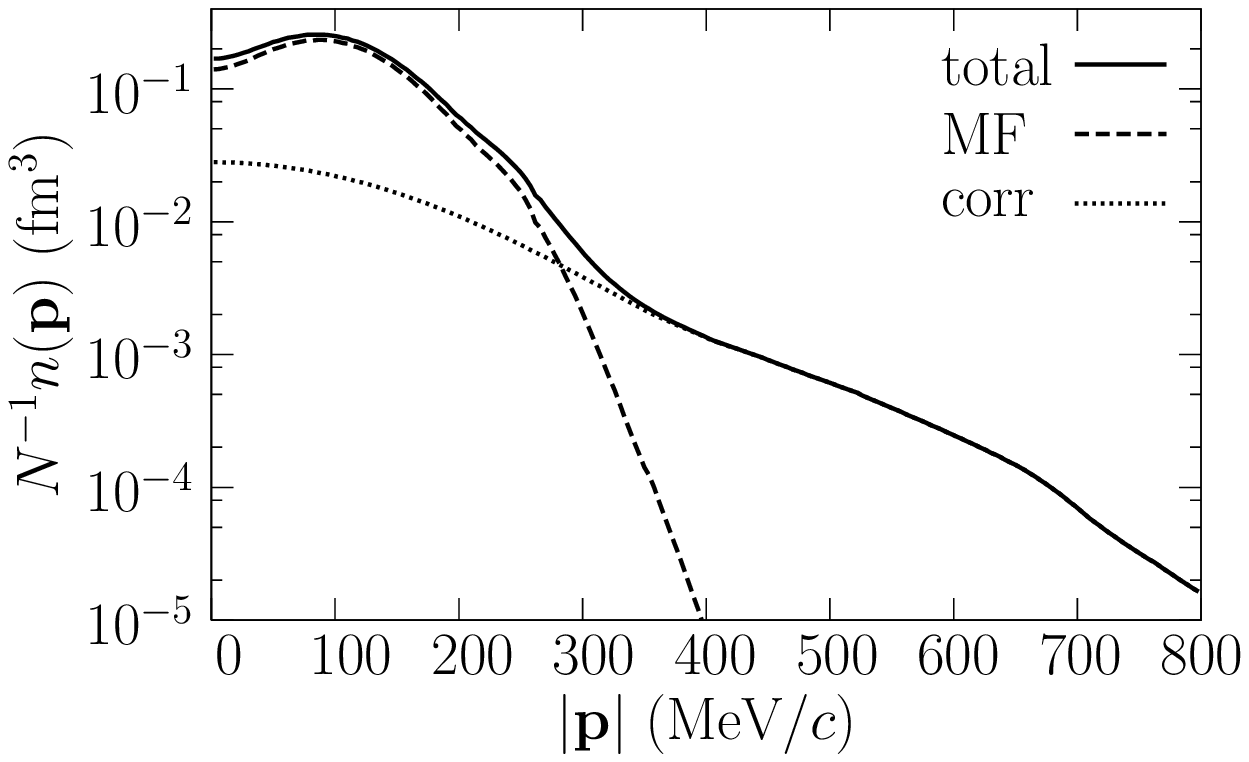}
\end{minipage}
\caption{Contributions to the momentum distribution of nucleons: the
mean field (dashed) and correlation part (dotted line) sum up to the
total momentum distribution (solid line). Left: Distribution
from~\cite{ref:Ciofi&Simula}. Right: Corresponding one calculated
from the Benhar's SF, by analogy divided into two parts (details in
text).}\label{fig:nContribs}
\end{figure}

To study how the difference between the momentum distributions
affects difference between the cross sections, we divided the
Benhar's distribution into MF and correlated part as
in~\Fref{fig:nContribs}: we kept the same value of
$n_\text{corr}(\ve p)$ for $\ve p=0$ and took care of its smooth
transition into the original distribution for high values of $\ve
p$. The calculated components of the total distribution were applied
to the simplest approximation of the spectral function (SSF).

\begin{figure}
\graphicspath{{plot/}}\centering
\begin{minipage}[l]{0.49\textwidth}%
\includegraphics*[scale=0.48]{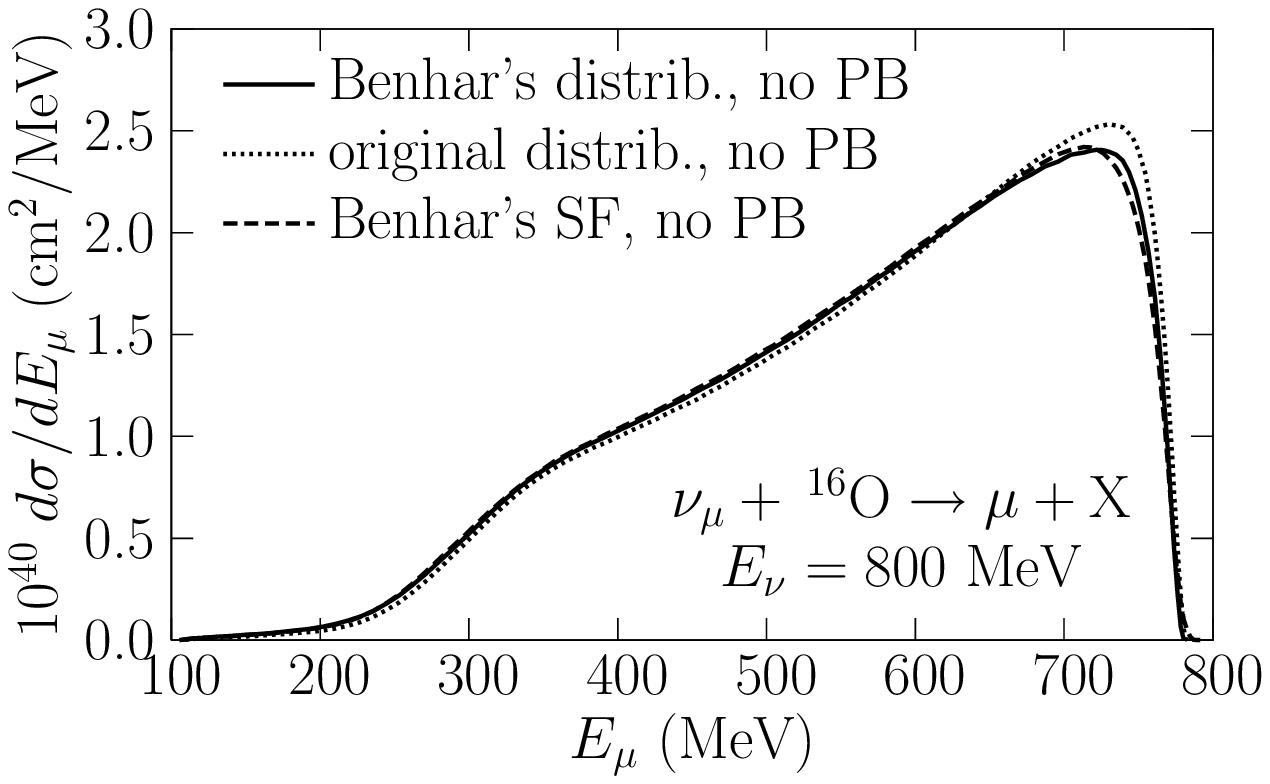}
\end{minipage}
\begin{minipage}[r]{0.49\textwidth}%
\includegraphics*[scale=0.48]{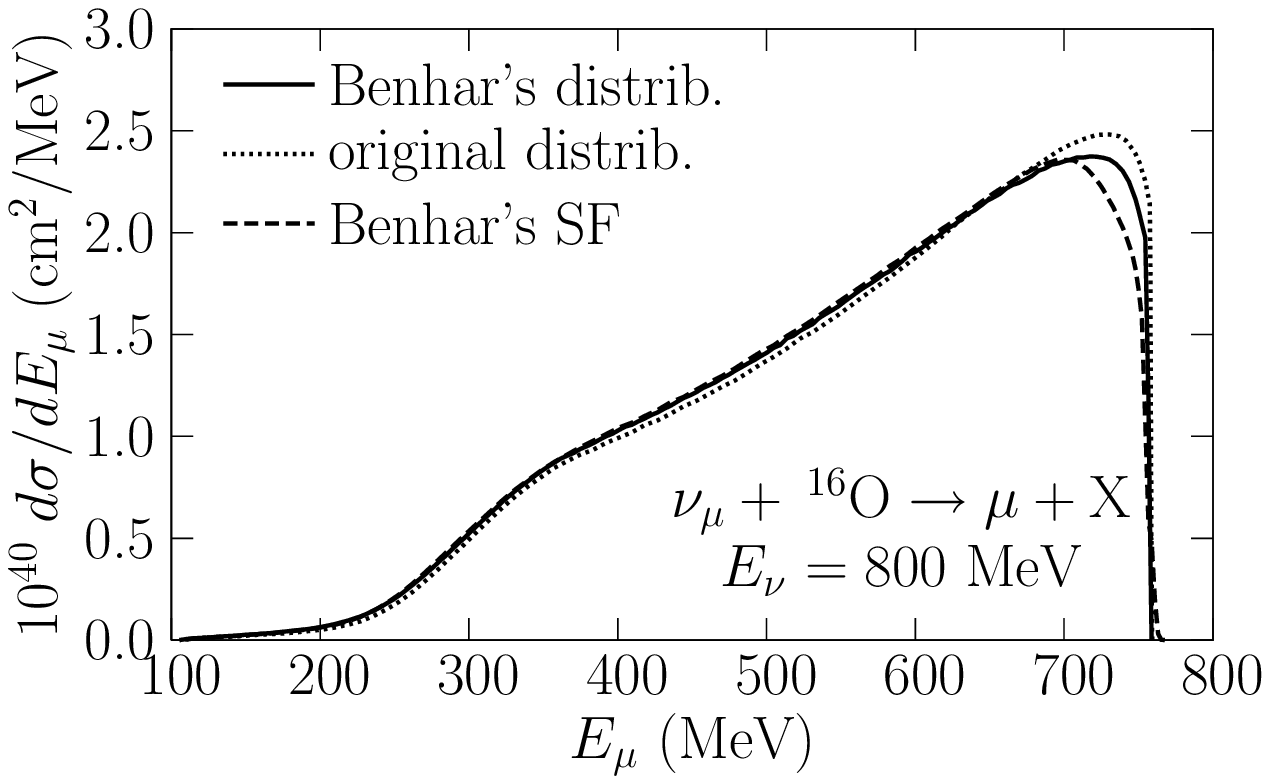}
\end{minipage}
\caption{Sensitivity of the quasielastic differential cross section
$d\sigma/dE_\mu$ on applied momentum distribution. The simplest
approximation with the momentum distribution used
previously~\cite{ref:Ankowski&Sobczyk}
(from~\cite{ref:Ciofi&Simula}; dotted line) and with ``the Benhar's
distribution'' (see~\fref{fig:nContribs}; solid line) compared to
the Benhar's spectral function (dashed line). Left: Without Pauli
blocking. Right: With Pauli blocking included.}\label{fig:cben}
\end{figure}

As it follows from \fref{fig:cben}, nearly whole discrepancy between
the cross section for SSF obtained in this way and the corresponding
one for the Benhar's SF disappeared when Pauli blocking (PB) is
absent. Inclusion of PB reveals different behavior of the two
approaches.

We conclude that the momentum distribution is responsible for the
height of the peak, but not for it's shape (compare solid and dotted
line in \fref{fig:cben}). Therefore the SSF diverges from the exact
SF when PB is present.

Another important information is the high sensitivity of
$d\sigma/dE_\mu$ on $n(\ve p)$. This feature is quite inconvenient,
because it suggests that one should take into account the difference
between momentum distribution of protons and neutrons, if
appropriate prediction of $d\sigma/dE_\mu$ for high $E_\mu$ is
required.



\section{Combination of the two effects}\label{sec:GSF}

\begin{figure}
\graphicspath{{plot/}}\centering
\begin{minipage}[l]{0.49\textwidth}%
\includegraphics*[scale=0.48]{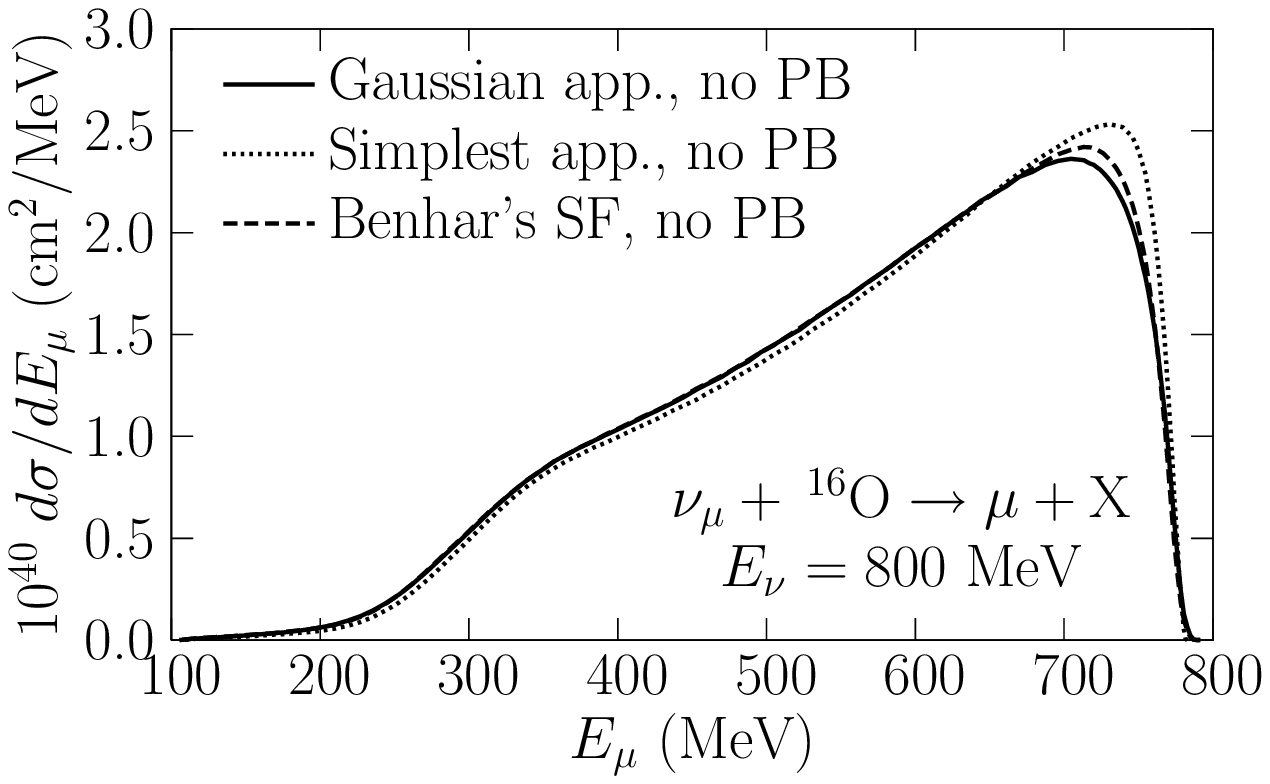}
\end{minipage}
\begin{minipage}[r]{0.49\textwidth}%
\includegraphics*[scale=0.48]{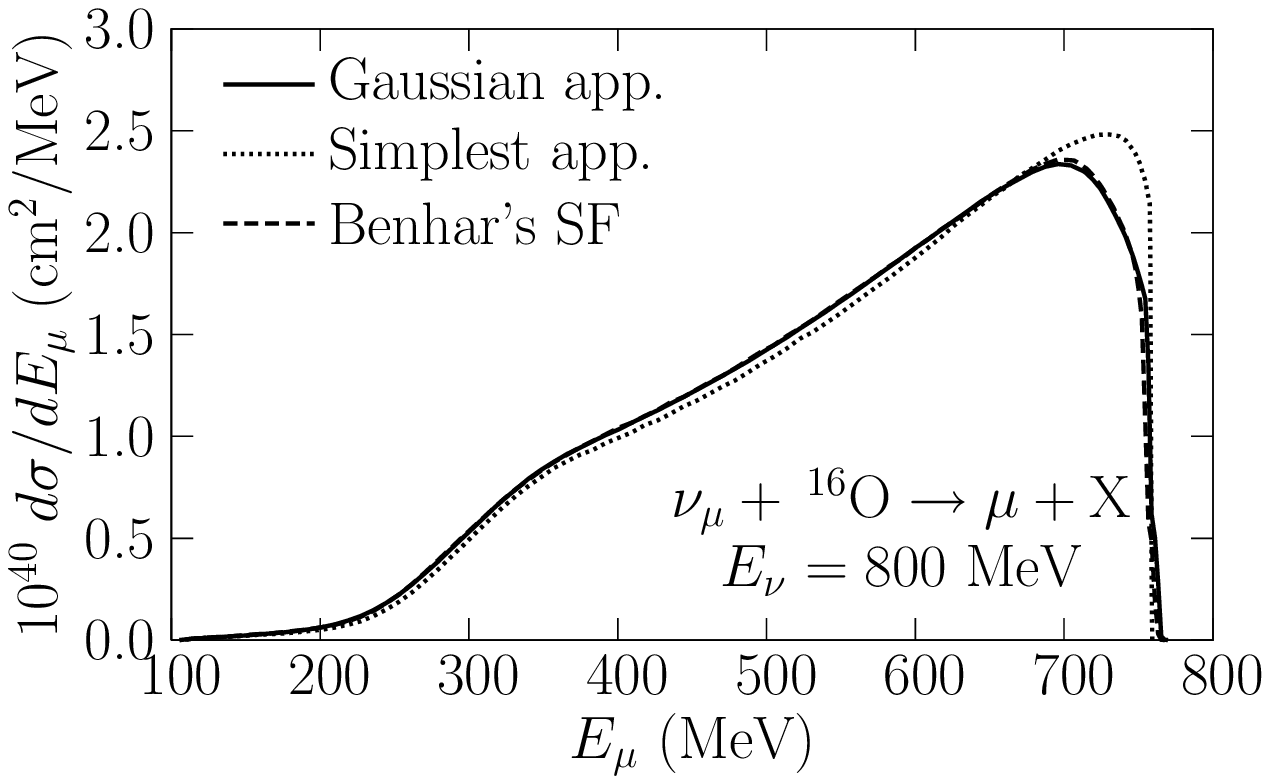}
\end{minipage}
\caption{Gaussian approximation with ``the Benhar's distribution''
(solid line) is significantly better then the simplest approximation
with the original distribution (dotted line) and nicely reproduces
results of the Benhar's spectral function (dashed line). Left:
Without Pauli blocking. Right: With Pauli blocking.}\label{fig:gsf}
\end{figure}

As we have seen in \sref{sec:MFTreatment}, the enhanced treatment of
the mean field SF affected the shape of the cross section's peak,
but it was a~bit too high, whereas the change of the momentum
distribution in \sref{sec:MomDistrib} reduced the hump's height,
however it's shape disagreed. When we combine the two improvements,
using both ``the Benhar's momentum distribution'' and the Gaussian
mean field SF, discrepancy vanishes almost completely, see
\fref{fig:gsf}.

We conclude that the remarks made in~\cite{ref:Ankowski&Sobczyk}
were right: the cross section of the Benhar's SF can be satisfactory
reproduced when one uses the same momentum distribution and the
refined treatment of the mean field SF.

Our approximation remains simple enough and can be applied to other
nuclei. The next goal will be $^{40}_{20}${Ca} where the needed
nuclear data is available.
\\

The author would like to express his gratitude to Jan T. Sobczyk for
stimulating discussions on the spectral function and to Omar Benhar
for providing his spectral function of oxygen.


\begin{thebibliography}{99}
\bibitem{ref:ICARUS}{
A.M.~Ankowski \etal (ICARUS Collaboration), {\tt arXiv:hep-ex/0606006}; %
J.~Kisiel, Nucl. Phys. {\bf B {\rm(Proc. Suppl.)} 155}, 205 (2006); %
F.~Arneodo \etal (ICARUS Collaboration), {\tt arXiv:hep-ex/0103008}.%
}
\bibitem{ref:T2K}{A.~Meregaglia, Nucl. Phys. {\bf B {\rm(Proc. Suppl.)} 155}, 248 (2006).}
\bibitem{ref:NuMI}{S.~Pordes, Nucl. Phys. {\bf B {\rm(Proc. Suppl.)} 155}, 225 (2006); D.~Finley \etal, Fermilab: {\bf FN-0776-E}.}
\bibitem{ref:Co_priv}{G.~Co', private communication.}
\bibitem{ref:Oset}{P.~Fern\'{a}ndez de C\'{o}rdoba \etal, {\tt arXiv:nucl-th/9612029}; E.~Marco and E.~Oset,  Nucl. Phys. {\bf A618}, 427 (1997).}
\bibitem{ref:Benhar&Fabrocini&Fantoni&Sick}{O.~Benhar \etal, Nucl. Phys. {\bf A579}, 493 (1994).}
\bibitem{ref:Benhar&Farina}{O. Benhar \etal, Phys. Rev.  {\bf D72}, 053005 (2005).}
\bibitem{ref:Benhar&Pandharipande}{O.~Benhar and V.R.~Pandharipande, Phys. Rev. {\bf C47}, 2218 (1993).}
\bibitem{ref:Ankowski&Sobczyk}{A.M.~Ankowski and J.T.~Sobczyk, {\tt arXiv:nucl-th/0512004}.}
\bibitem{ref:Frullani&Mougey}{S.~Frullani and J.~Mougey, Adv. Nucl. Phys. {\bf 14}, 1 (1984).}
\bibitem{ref:Benhar&Fabrocini&Fantoni}{O.~Benhar, A.~Fabrocini, and S.~Fantoni, Nucl. Phys. {\bf A505}, 267 (1989).}
\bibitem{ref:Ciofi&Simula}{C.~Ciofi degli Atti and S.~Simula, Phys. Rev. {\bf C53}, 1689 (1996).}
\bibitem{ref:Ciofi&Liuti&Simula}{C.~Ciofi degli Atti, S.~Liuti, and S.~Simula, Phys. Rev. {\bf C41}, R2474 (1990).}
\bibitem{ref:Kulagin&Petti}{S.A.~Kulagin and R.~Petti, Nucl. Phys. {\bf A765}, 126 (2006).}
\bibitem{ref:OccProbs}{G.A.~Lalazissis, S.E.~Massen, and C.P.~Panos, Phys. Rev. {\bf C48}, 944 (1993).}
\bibitem{ref:CaData}{V.V.~Johnson, C.~Mahaux, Phys. Rev. {\bf C38}, 2589 (1988).}
\end{thebibliography}
\end{document}